\documentclass[showpacs,prb,twocolumn,floats]{revtex4-1}
\usepackage{amsfonts}

\usepackage{amsmath}
\usepackage{amssymb}

\usepackage{graphicx}

\begin{document}

\title{Four band tunneling in bilayer graphene}
\date{\today }
\author{B. Van Duppen}
\email{ben.vanduppen@ua.ac.be}
\author{F. M. Peeters}
\email{francois.peeters@ua.ac.be}
\affiliation{Department of Physics, University of Antwerp, Groenenborgerlaan 171, B-2020 Antwerp, Belgium}

\begin{abstract}
The conductance, the transmission and the reflection probabilities through rectangular potential barriers and pn-junctions are obtained for bilayer graphene taking into account the four bands of the energy spectrum. We have evaluated the importance of the skew hopping parameters $\gamma_3$ and $\gamma_4$ to these properties and show that for energies $E>\gamma_1/100$ their effect is negligible. For high energies two modes of propagation exist and we investigate scattering between these modes. For perpendicular incidence both propagation modes are decoupled and scattering between them is forbidden. This extends the concept of pseudospin as defined within the two band approximation to a four band model and corresponds to the (anti)symmetry of the wavefunctions under in-plane mirroring. New transmission resonances are found that appear as sharp peaks in the conductance which are absent in the two band approximation. The application of an interlayer bias to the system: 1) breaks the pseudospin structure, 2) opens a bandgap that results in a distinct feature of suppressed transmission in the conductance, and 3) breaks the angular symmetry with respect to normal incidence in the transmission and reflection.
\end{abstract}

\maketitle

\section{Introduction}

Bilayer graphene (BLG) is a system consisting of two Bernal\cite{Bernal1924} stacked graphene monolayers\cite{McCann2006,Novoselov2006}. Whereas monolayer graphene has a linear electronic spectrum in the vicinity of each corner of the Brillouin zone ($\mathbf{K}$ and $\mathbf{K^{\prime }}$ points, also known as \textit{Dirac points}), bilayer graphene has four hyperbolic bands. Two of these bands touch in the $K$ point at zero energy making BLG a gapless semiconductor, the other two bands are displaced by an energy of $\gamma _{1}=377meV$ with respect to the touching bands\cite{Partoens2006}. The application of a potential that breaks the interlayer symmetry can however create a tunable bandgap\cite{McCann2006d, Zhang2009, Castro2007}. Often, a low energy approximation\cite{McCann2006} is made that is valid for electron kinetic energy much smaller than the interlayer hopping parameter $\gamma _{1}$. This so called two band approximation has a quadratic dispersion and is only valid near the Dirac point.

In monolayer graphene Klein tunneling results in a 100\% probability for perpendicular transmission through potential barriers, as predicted\cite{Katsnelson2006} and observed experimentally\cite{Young2009,Stander2009}. For bilayer graphene due to the conservation of pseudospin, no Klein tunneling is expected and this was confirmed theoretically within the two band approximation\cite{Katsnelson2006,Barbier2010}. In this case there are electronic states available that are not accessible for penetration into the potential barrier, which was called a \textit{cloaking} of those states\cite{Gu2011}.

Previous work was based on the two band approximation which we extend here to the four band model. This allows us to investigate the electronic properties at higher Fermi level, i.e. beyond $\gamma _{1}$, and for a higher electrostatic potential. Recent experimental progress has allowed to access this energy region and measurements of the electronic transport in this region is expected\cite{Efetov2010, Ye2011, Efetov2011}. We calculate the transmission and reflection probabilities within the same band and between the two bands for electrons impinging on a rectangular potential barrier (pnp-junction) and a potential step (pn-junction) at different angles of incidence and investigate the effect of the application of an interlayer bias to the transport properties. Furthermore we compare the energy dependence of the conductance calculated within the two band and the four band model and point out major differences.

Our findings show that at low energy and low potential the same phenomena occur as those predicted by the two band model and that the existence of pseudospin is related to the (anti)symmetry of the wavefunctions. This relation is a consequence of the symmetry of the crystal and is therefore also valid when the skew hopping parameters are taken into account. Outside the range of validity of the two band approximation however, we predict several new phenomena: 1) for high potential barriers new resonances are found that are absent in the two band model; 2) the use of four bands introduces a new mode of propagation; 3) we investigate a new form of cloaking and calculate the scattering between the two modes of propagation; and 4) the newly discovered resonances and the second mode of propagation lead to distinctive features in the conductance that are absent in the two band calculation. We also justify the use of only the nearest neighbor interlayer hopping parameters. Finally, we show that the application of an interlayer bias not only opens a bandgap and therefore suppresses the transmission in this region, it also unexpectedly breaks the angular symmetry with respect to normal incidence when only one valley is considered.

The paper at hand is organized as follows. In Sec. \ref{sec:PropModes} we present the formalism, indicate the different propagating modes, define the eight different transmission and reflection probabilities for the four band  model and explain the transition from the four band to the two band model at low energy. The effect of the skew hopping parameters on our results is critically examined. In Sec. \ref{TimeSymmetry} we analyze the symmetries of the system to explain the transmission for normal incidence and the surprising occurrence of angular asymmetry at non normal incidence The numerical results for the conductance, transmission and reflection at non normal incidence for pn-junctions and potential barriers with and without interlayer bias are discussed in Sec. \ref{NumResults}, and we show the effect of the skew hopping parameters on the transmission. In Sec. \ref{Conclusion} we summarize the main points of this paper.

\section{The propagating modes}\label{sec:PropModes}

In this section we discuss the dispersion relation of BLG and the resulting propagating modes. It turns out that electrons in BLG can propagate via two different modes and we find that when electrons impinge perpendicular on a potential barrier, it is not possible to scatter between those modes.

\begin{figure}[t]
\centering
\includegraphics[width= 8cm]{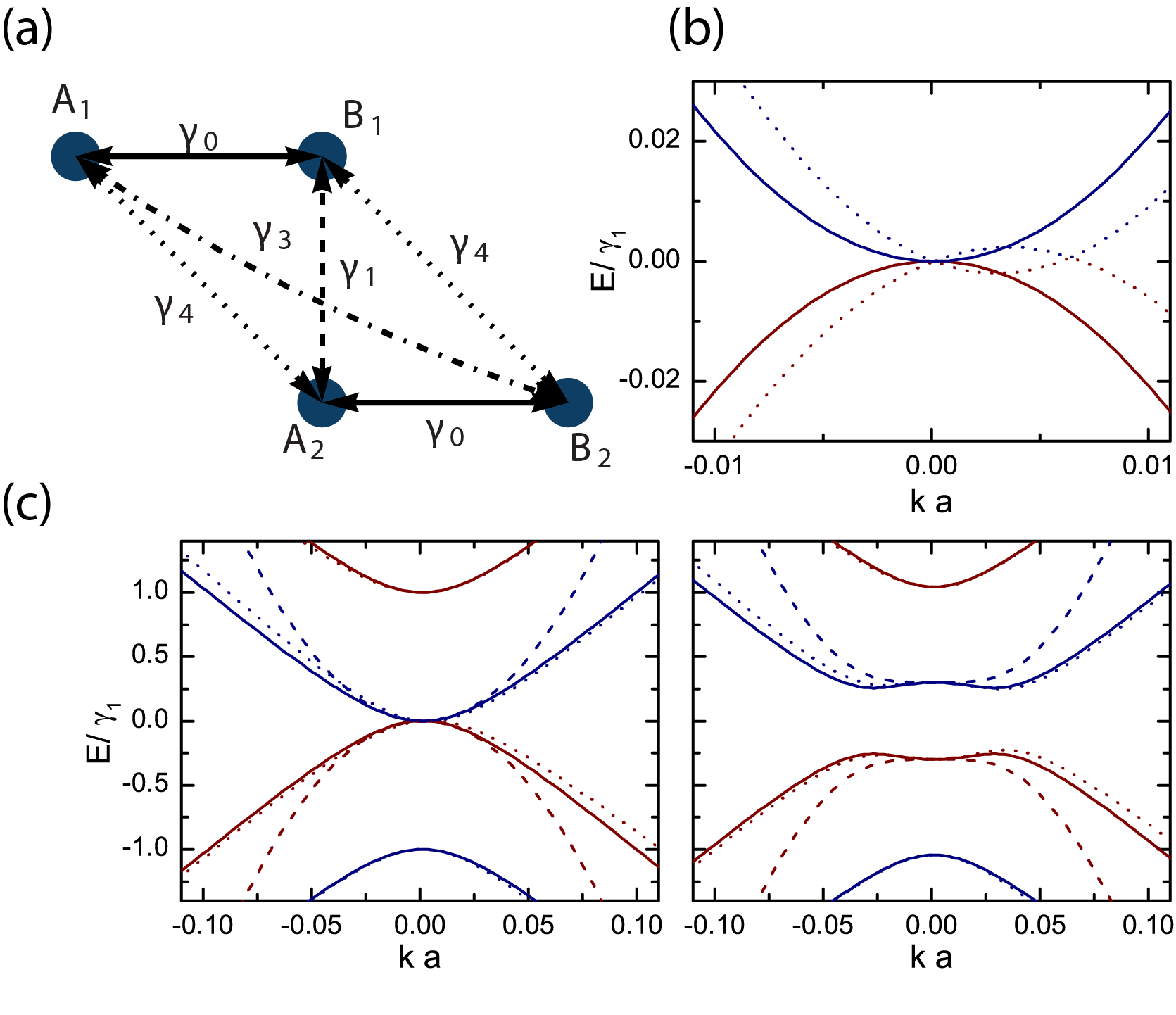}
\caption{(Colour online) (a) Schematic presentation of the sublattices in bilayer graphene. The arrows indicate the different interatomic hopping parameters. (b) Energy spectrum of bilayer graphene near one of the
Dirac points for low energy. The dotted curve corresponds to the spectrum accounting for all interatomic hopping parameters, the solid curve accounts only for $\gamma_0$ and $\gamma_1$. (c) Total energy spectrum of bilayer graphene for (left) an unbiased system and (right) a system with interlayer bias $\delta=0.3 \gamma_1$. The dotted curves also account for the skew hopping parameters while the solid curve considers only nearest neighbor interlayer hopping. The dashed curve corresponds to the spectrum of the two band approximation. All bands in (b) and (c) are coloured according to their relation to the (anti)symmetric states. }
\label{Bands}
\end{figure}

We model the BLG crystal as two hexagonal monolayer flakes with in plane interatomic distance\cite{Partoens2007} $a=0.142$nm, each consisting of two nonequivalent sublattices with atoms $A_{1}$ and $B_{1}$ for the top layer and $A_{2}$ and $B_{2}$ for the bottom one. These two layers are stacked according to Bernal stacking which places the $A_{2}$ atom just above the $B_{1}$ as schematically shown in Fig. \ref{Bands}(a). In both layers, each $A$ atom is surrounded by three $B$ atoms and vice versa. The intralayer coupling between these atoms is $\gamma_{0}\approx 3eV$. Between the $A_{2}$ and $B_{1}$ atoms the interlayer coupling is $\gamma _{1}\approx 0.4eV$ while the skew hopping energy between the other two sublattices are denoted as $\gamma _{3}\approx 0.3eV$ and $\gamma _{4}\approx 0.1 eV$. These interatomic coupling parameters are depicted in Fig. \ref{Bands}(a). The contribution of the skew hopping parameter $\gamma_3$ results in the so called\cite{McCann2006, McCann2007} \textit{trigonal warping}, an effect occurring only at very low energy ($E\lesssim 4meV$). The other parameter, $\gamma_4$ has an even lower impact on the electronic properties as will be discussed. Therefore, we will often neglect these two $\gamma$-parameters allowing for a more comprehensive discussion.

Following the continuum nearest-neighbor tight-binding formalism, the effective Hamiltonian near the \textbf{K} point and the corresponding eigenstates are given by\cite{Snyman2007,Neto2009}
\begin{equation}
H_{4}=%
\begin{bmatrix}
V+\delta  & v_{F}\pi ^{\dag } & -v_{4}\pi ^{\dag } & v_{3}\pi  \\
v_{F}\pi  & V+\delta  & \gamma _{1} & -v_{4}\pi ^{\dag } \\
-v_{4}\pi  & \gamma _{1} & V-\delta  & v_{F}\pi ^{\dag } \\
v_{3}\pi ^{\dag } & -v_{4}\pi  & v_{F}\pi  & V-\delta
\end{bmatrix},
\Psi _{4}=\left(
\begin{array}{c}
\psi _{A_{1}} \\
\psi _{B_{1}} \\
\psi _{A_{2}} \\
\psi _{B_{2}}%
\end{array}%
\right) .  \label{StartHam}
\end{equation}%
Here, $v_{F}=\frac{\gamma _{0}}{\hbar }\frac{3a}{2}\approx 10^{6}m/s$ is the Fermi velocity for electrons in monolayer graphene and $v_{3,4}=v_F \gamma_{3,4}/\gamma_0$ are related to the skew hopping parameters. $\mathbf{\pi }=p_{x}+ip_{y}=\hbar (k_{x}+ik_{y})$ is the in-plane momentum relative to the Dirac point, $V$ is a general potential term and $\delta$ corresponds to an externally induced interlayer potential difference. Due to the dimensionality of the Hamiltonian, the eigenstate of the system is a four component spinor. Neglecting the skew hopping parameters, the energy spectrum of this Hamiltonian is given by
\begin{equation}
\varepsilon =l\left( \sqrt{k^{2}+\Delta ^{2}+\frac{\Gamma _{1}^{2}}{2}\pm \sqrt{k^{2}\left( \Gamma _{1}^{2}+4\Delta ^{2}\right) +\frac{\Gamma _{1}^{2}}{2}}}\right) ,
\end{equation}
where $l=\pm 1$ and $k=\sqrt{k_{x}^{2}+k_{y}^{2}}$. We have used the reduced variables $\Gamma_1 =\gamma_{1}/\hbar v_{F}$, $\Delta =\delta/\hbar v_{F}$ and $\varepsilon =(E-V)/\hbar v_{F}$ with $E$ the energy of the electrons. For a system without interlayer potential bias $\delta$, this result reduces to the one previously found by Snyman \textit{et al.}\cite{Snyman2007}. The energy spectra corresponding to these systems are displayed in Fig. \ref{Bands}(c). In Fig. \ref{Bands}(b) we show the dispersion relation when taking also account of the skew hopping parameters $\gamma_3$ and $\gamma_4$. Their effect is clearly negligible for $E>\gamma_1/100 \approx 4meV$.

Vice versa it is possible to calculate the value of $k_{x}$ as a function of the energy and $k_{y}$. This corresponds to the wave vectors of the plane wave solutions of the Schr\"{o}dinger equation $H\Psi =E\Psi $ and is given by
\begin{equation}
k^{l}=\sqrt{\varepsilon ^{2}+\Delta ^{2}+l\sqrt{\varepsilon ^{2}\left(\Gamma _{1}^{2}+4\Delta ^{2}\right) -\Gamma _{1}^{2}\Delta ^{2}}-k_{y}^{2}},
\end{equation}
which for $\Delta=0$ reduces to
\begin{equation}
k^{l}=\sqrt{\varepsilon ^{2}+l\varepsilon \Gamma _{1}-k_{y}^{2}}.
\label{KvectorEq}
\end{equation}
Depending on the value of $\varepsilon $ relative to $\Gamma_1 $, $\Delta$ and $k_{y} $ this wave vector can be real or imaginary. This means that it can represent a traveling or an evanescent plane wave. Without the interlayer bias, when $0<\varepsilon <\Gamma_1 $, the $k^{-}$ is imaginary while $k^{+}$ is real and then propagation is only possible using the $k^{+}$ channel. When $\varepsilon >\Gamma_1 $ however, $k^{-}$ becomes real too, providing a new mode for propagation. For $\varepsilon <0$, a similar argument leads to propagation via $k^{-}$ when $|\varepsilon |<\Gamma_1 $ and two ways of propagation when $|\varepsilon |>\Gamma_{1}$.

One obtains the two band approximation for $\varepsilon \ll\Gamma_{1}$ with the condition that $\Delta$ and $\varepsilon$ are of the same order of magnitude. Therefore, one can neglect these terms up to second order in Eq. (\ref{KvectorEq}), resulting in the wave vector
\begin{equation}
k^{l} \approx \sqrt{l\Gamma _{1}\sqrt{\varepsilon ^{2}-\Delta ^{2}}-k_{y}^{2}},
\end{equation}
which for $\Delta=0$ reduces to
\begin{equation}
k^{l} \approx \sqrt{l\varepsilon \Gamma _{1}-k_{y}^{2}},
\label{TwoBandWavevector}
\end{equation}
with the energy spectrum
\begin{equation}
\varepsilon  \approx \sqrt{\left[ \frac{l}{\Gamma _{1}}\left( \left( k^{l}\right)^{2}+k_{y}^{2}\right) \right] ^{2}+\Delta ^{2}},
\end{equation}
and for $\Delta=0$
\begin{equation}
\varepsilon \approx \frac{l}{\Gamma _{1}}\left( \left( k^{l}\right) ^{2}+k_{y}^{2}\right).
\end{equation}
This spectrum is superimposed in Fig. \ref{Bands}(c) as dashed curves. It agrees with the full spectrum only near the Dirac point. The validity of the approximation is based on the increase in energy near the atomic sites of the $B_{1}$ and $A_{2}$ atoms which influence each other. For low Fermi energy, it therefore makes sense to take into account only the orbital wave functions near the other two atoms. This reduces the $4\times 4$ Hamiltonian, Eq. $\left( \ref{StartHam}\right) $, and replaces it with the approximate one given by\cite{McCann2006}
\begin{equation}
H_{2}=-\frac{\hbar ^{2}v_{F}^{2}}{\gamma _{1}}\left[
\begin{array}{cc}
V^{\prime \prime} + \Delta^{\prime \prime}  & \left( k_{x}-ik_{y}\right) ^{2} \\
\left( k_{x}+ik_{y}\right) ^{2} & V^{\prime \prime} - \Delta^{\prime \prime}
\end{array}%
\right] ,\Psi =\left(
\begin{array}{c}
\psi _{A_{1}} \\
\psi _{B_{2}}%
\end{array}%
\right) ,  \label{TwoHam}
\end{equation}%
where $V^{\prime \prime }=-\gamma _{1}V/\left( \hbar ^{2}v_{F}^{2}\right) $ and $\Delta^{\prime \prime} = -\gamma_1 \delta/\left( \hbar ^{2}v_{F}^{2}\right)$. The two spinor plane wave solution of the Schr\"{o}dinger equation of this Hamiltonian consists of a propagating wave with wave vector given by Eq. (\ref{TwoBandWavevector}) and an evanescent mode with inverse decay length
\begin{eqnarray}
\kappa &=&\sqrt{\sqrt{\varepsilon^2-\Delta^2} \Gamma_{1}+k_{y}^{2}},\\
&=&\sqrt{\varepsilon \Gamma_{1}+k_{y}^{2}} \text{ for }\Delta =0,
\end{eqnarray}%
which corresponds to the imaginary part of $k^l$ with $l=-1$ in Eq. (\ref{TwoBandWavevector}). There is no positive energy value that can make this quantity imaginary and so it can only represent a traveling state when $\varepsilon <0$ which corresponds to a hole state. In contrast to the four band treatment, there is no second mode of propagation.
The introduction of an interlayer bias term $\delta$ in the system has a strong influence on the electronic properties. As was pointed out earlier\cite{McCann2006d}, it opens a gap in the spectrum which completely changes the low energy behavior of the electrons. When an interlayer bias is applied, the effect of the skew hopping parameters is of even less importance for the spectrum and as shown in Fig. \ref{Bands}(c). Notice that the two band approximation fails to describe the spectrum of the system accurately.

\begin{figure}[th]
\begin{center}
\includegraphics[width=0.45 \textwidth]{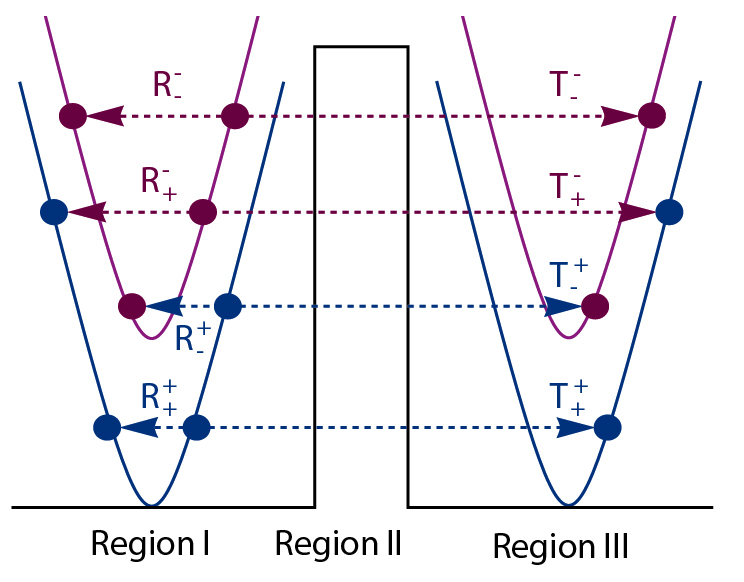}
\end{center}
\caption{(Colour online) Schematic representation of the transmission and reflection probabilities for a rectangular potential barrier.}
\label{Channels}
\end{figure}

The potential we will consider is similar to the one used by Katsnelson \textit{et al.}\cite{Katsnelson2006}, but now with the addition of an interlayer potential bias term. It consists of a one dimensional potential barrier of width $d$ given by
\begin{equation}
V(x)=%
\begin{cases}
\begin{array}{cc}
0 & \text{ if } x<0 \\
V_0 + \xi \delta & \text{ if }0\leq x\leq d \\
0 & \text{ if } x>d%
\end{array}
&
\begin{array}{c}
\text{ (region I)} \\
\text{ (region II)} \\
\text{ (region III)}%
\end{array}%
\end{cases}%
\end{equation}%
where $\xi = +1$ for the first layer and $\xi=-1$ for the second layer. This barrier is shown in Fig. \ref{Channels} and because it is translational invariant in the $y$ direction, $k_{y}$ is a conserved quantity. Using the Hamiltonians given by Eqs. $\left( \ref{StartHam}\right) $ and $\left( \ref{TwoHam}\right) $, we can now calculate the transmission and reflection probabilities for electrons impinging on the barrier. The two band model only allows for one mode of propagation, leading to one transmission $(T)$ and one reflection $(R)$ channel. For sufficiently large energy however, in the four band model, it is possible to propagate through two distinct modes. Therefore, we will have four reflection and four transmission channels. For the transmission these are two non scattered channels, which we denote as $T_{+}^{+}$ and $T_{-}^{-}$ for propagation via $k^{+}$ and $k^{-}$ respectively and two scattered channels in which the particle enters through one channel and exits through another one. We will denote them as $T_{-}^{+}$ for scattering from the $k^{+}$ band to the $k^{-}$band and $T_{+}^{-}$ for the other direction. A similar definition holds for the $R_{\pm }^{\pm }$ reflection channels. The eight channels are schematically depicted in Fig. \ref{Channels}.

Using the transmission probabilities, we can calculate the conductance as function of the energy given by the Landauer-B\"{u}ttiker formula \cite{Blanter2000}
\begin{equation}
G\left( E\right) =G_{0}\frac{W}{2\pi }\int_{-\infty }^{\infty
}dk_{y}\sum_{l,m=\pm }T_{m}^{l}\left( E,k_{y}\right) ,
\end{equation}%
where $G_{0}=4e^{2}/h$, which is four times the quantum of conductance due to spin and valley degeneracy and $W$ is the width of the sample in the $y$ direction.

\section{Transmission probabilities and symmetries\label{TimeSymmetry}}

\begin{figure}[tb]
\begin{center}
\includegraphics[width=0.35 \textwidth]{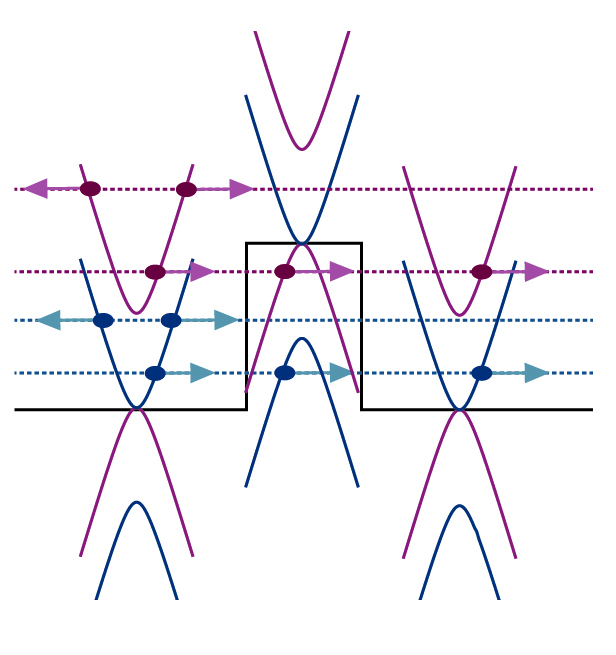}
\end{center}
\caption{(Colour online) Schematic representation of the energy spectrum inside and outside the barrier region. The dots and arrows indicate the energy regions for which electrons impinging perpendicularly on the barrier will be transmitted or reflected.}
\label{TransmissionUA}
\end{figure}

Before calculating the transmission probability by matching the components of the spinor wave function at the boundary as explained by Barbier \textit{et al. }\cite{Barbier2010} and Snyman \textit{et al. }\cite{Snyman2007}, a simplification can be made by transforming the Hamiltonian from Eq. $\left( \ref{StartHam}\right) $. Constructing symmetric and antisymmetric combinations of the spinor components by combining the atomic wave functions $\psi_{A_{1}} $ with $\psi _{B_{2}}$ and $\psi _{B_{1}}$ with $\psi _{A_{2}}$, the Hamiltonian is transformed into
\begin{equation}
H_{4} =\hbar v_{F}%
\begin{bmatrix}
V^{\prime }-v_{3}^{\prime }k_{x} & u_{4}^{+} k_{x} & iv_{3}^{\prime }k_{y}+\Delta & -iu_{4}^{-}k_{y} \\
u_{4}^{+} k_{x} & V^{\prime }-\Gamma _{1} & iu_{4}^{+} k_{y} & \Delta  \\
-iv_{3}^{\prime }k_{y}+\Delta  & -iu_{4}^{+}k_{y} & V^{\prime }+v_{3}^{\prime }k_{x} & u_{4}^{-}k_{x} \\
iu_{4}^{-} k_{y} & \Delta  & u_{4}^{-} k_{x} & V^{\prime }+\Gamma _{1}
\end{bmatrix}
,  \label{eq:HamTransfo}
\end{equation}
with the new spinor
\begin{equation}
\Psi _{4}^{\prime } =\frac{1}{\sqrt{2}}\left(
\begin{array}{c}
\psi _{A_{1}}-\psi _{B_{2}} \\
\psi _{B_{1}}-\psi _{A_{2}} \\
\psi _{A_{1}}+\psi _{B_{2}} \\
\psi _{A_{2}}+\psi _{B_{1}}
\end{array}
\right) =\left(
\begin{array}{c}
\Psi _{4,+} \\
\Psi _{4,-}
\end{array}
\right) ,
\end{equation}
where we have introduced the reduced potential $V^{\prime }=V/\hbar v_{F}$ and $u_{4}^{\pm}=\left(1 \pm v_4^{\prime}\right)$ with the dimensionless skew hopping velocities $v^{\prime}_{3,4}=v_{3,4}/v_F=\gamma_{3,4}/\gamma_0$ which turn out to be very small, i.e. $v^{\prime}_3 \approx 0.09$ and $v^{\prime}_4 \approx 0.04$. These small values advocate the neglect of the skew hopping parameters in the discussion a little bit further.
The four component spinor can be seen as a combination of two two-spinors $\Psi _{4,+}$ and $\Psi _{4,-}$ which are respectively antisymmetric and symmetric with respect to the exchange $A_{1}\leftrightarrow B_{2}$ and $A_{2}\leftrightarrow B_{1}$. This exchange corresponds to a reflection of the system by an in-plane mirror. Notice that for normal incidence and for an unbiased system, i. e. when $k_{y}=0$ and $\Delta=0$, the Hamiltonian is block diagonal in this basis. This means that it represents two non interacting one dimensional systems with eigenfunctions $\Psi _{4,\pm }$ which are described by the $2\times 2$ Hamiltonian
\begin{equation}
H_{4,l}=\hbar v_{F}%
\begin{bmatrix}
V^{\prime }-lv_{3}^{\prime }k_{x} & \left( 1+v_{4}^{\prime }\right) k_{x} \\
\left( 1+v_{4}^{\prime }\right) k_{x} & V^{\prime }-l\Gamma _{1}%
\end{bmatrix}.
\end{equation}
Neglecting the skew hopping terms, this Hamiltonian describes a one dimensional monolayer of graphene with a potential term that breaks the sublattice symmetry and corresponds to a Dirac Hamiltonian with a mass term\cite{Matulis2009}. By calculating the energy spectrum of this Hamiltonian, we can relate $\Psi _{4,l}$ to the bands corresponding to $k^{l}$. In Fig. \ref{TransmissionUA} the four energy bands are depicted inside and outside the barrier region. Bands belonging to the same $\Psi _{4,l}$ have the same colour. From this we can see that within different energy ranges the transmission at normal incidence depends on the availability of states corresponding to the same $l$.

In the two band approximation, and neglecting the skew hopping parameters, a similar symmetry transformation leads to the Hamiltonian
\begin{equation}
H_{2}=-\frac{\hbar ^{2}v_{F}^{2}}{\gamma _{1}}\left[
\begin{array}{cc}
k_{x}^{2}-k_{y}^{2}+V^{\prime \prime } & 2ik_{x}k_{y} + \Delta^{\prime \prime} \\
-2ik_{x}k_{y} + \Delta^{\prime \prime} & k_{y}^{2}-k_{x}^{2}+V^{\prime \prime }%
\end{array}%
\right] .
\end{equation}%
For normal incidence and without bias, the system corresponds to that of two non interacting Schr\"{o}dinger particles described by the Hamiltonian
\begin{equation}
H_{2,l}=-l\frac{\hbar ^{2}}{2m}k_{x}^{2}+V,
\end{equation}
and the wave functions
\begin{equation}
\Psi _{2,l}=\frac{1}{%
\sqrt{2}}\left( \psi _{A_{1}}+l\psi _{B_{2}}\right),
\end{equation}%
where $l=\pm 1$ and $m=\gamma _{1}/2v_{F}^{2}$. Since the system is now described by a two component spinor $\Psi _{2}$, one can extend the spinorial analogy of the charge carriers in monolayer graphene by introducing a \textit{pseudospin}\cite{Katsnelson2006}. The value of $l$ now corresponds to the pseudospin state of the particle. As shown above the pseudospin is the consequence of the (anti)symmetric nature of the wave functions under in-plane mirroring. Although the analogy with the normal spin properties of an electron are more pronounced within the two band model, the four band model incorporates the same symmetry and therefore the notion of pseudospin is applicable in this model as the value of $l$. Note that these symmetry considerations also hold when the skew hopping parameters are included because the Hamiltonian of Eq. (\ref{eq:HamTransfo}) is block diagonal at normal incidence for a system without bias.

The different energy ranges in which at normal incidence and for $\delta=0$ new phenomena are expected are defined by the value of $\varepsilon =(E-V)/\hbar v_{F}$ and thus by the difference between the kinetic energy and the height of the potential in the region under consideration. When $V^{\prime }-\Gamma_{1}<\varepsilon _{I}<\Gamma_{1}$ outside the barrier region, but $\varepsilon _{II}<0$ inside, for positive $\varepsilon _{I}$ the $\Psi _{+}$ state which propagates outside will not be able to propagate inside the barrier. The absence of propagating $\Psi _{+}$ states inside the barrier suppresses the transmission in this energy region. The fact that the transmission is suppressed even though there are propagating $\Psi _{-}$ states inside the barrier has been noticed before and was called a \textit{cloaking} of the $\Psi _{-}$ states\cite{Gu2011} and is present in both models.

When $\varepsilon _{I}>\Gamma_{1}$ outside the barrier, the $\Psi _{-}$ states can also propagate in regions I and III. Inside the barrier however, these states are trapped and their energy spectrum is discrete. This means that when electrons propagate through the $\Psi _{-}$ state, they can propagate inside the barrier region only if the energy matches one of the discrete energy levels inside. The discretisation condition corresponds to Fabry-Perot resonances, namely $d=n\lambda/2$ where $\lambda =2\pi /k^{\pm }$. This results in resonances whose energies are, in the four band model without skew hopping or interlayer bias, given by\cite{Snyman2007}
\begin{equation}
\varepsilon _{II}\left( n,k_{y}\right) =-\frac{l\Gamma_{1}}{2}\pm \sqrt{\left(
\frac{n\pi }{d}\right) ^{2}+k_{y}^{2}+\frac{1}{4}\Gamma_{1}^{2}}.
\label{eq:Resonances}
\end{equation}%
The above reasoning is valid for (near) normal incidence. The Hamiltonian of Eq. $\left( \ref{eq:HamTransfo}\right) $ however mixes up the two states for non-normal incidence.

To calculate the transmission probabilities at non-normal incidence, the transfer matrix method together with appropriate boundary conditions was implemented. The plane wave solution for the Schr\"{o}dinger equation of the four band model is a four component spinor and can be represented by a product of matrices
\begin{equation}
\Psi =\mathcal{PEC},
\label{eq:MatrixSpinor}
\end{equation}%
in which $\mathcal{E}$ corresponds to a $4\times 4$ diagonal matrix consisting of exponential terms and $\mathcal{P}$ is for $\delta=0$ given by
\begin{equation}
\mathcal{P}=\left[
\begin{array}{cccc}
1 & 1 & 0 & 0 \\
\frac{k^{+}}{\varepsilon } & -\frac{k^{+}}{\varepsilon } & -\frac{ik_{y}}{%
\varepsilon } & -\frac{ik_{y}}{\varepsilon } \\
0 & 0 & 1 & 1 \\
-\frac{ik_{y}}{\varepsilon } & -\frac{ik_{y}}{\varepsilon } & \frac{k^{-}}{%
\varepsilon } & -\frac{k^{-}}{\varepsilon }%
\end{array}%
\right] ,
\end{equation}%
in which $k^\pm$ is defined in Eq. (\ref{KvectorEq}). To find the transmission and reflection probabilities, one has to equate the wave functions at the borders of the potential barrier. This results in two times four equations from which we obtain the components of the vector $\mathcal{C}$. Considering the boundary conditions of the system, this vector is given by
\begin{equation}
\mathcal{C}_{I}^{l}=\left(
\begin{array}{c}
\delta _{l,1} \\
r_{+}^{l} \\
\delta _{l,-1} \\
r_{-}^{l}%
\end{array}%
\right) ,\text{ }\mathcal{C}_{III}^{l}=\left(
\begin{array}{c}
t_{+}^{l} \\
0 \\
r_{-}^{l} \\
0%
\end{array}%
\right) ,
\end{equation}%
where $l$ indicates the wave vector $k^{\pm }$ and $\delta _{l,\pm 1}$ is the Kronecker delta. Using the transfer matrix method finding the coefficients in these vectors corresponds to solving the matrix equation
\begin{equation}
\mathcal{C}_{I}^{l}=M_{I\rightarrow II}M_{II\rightarrow III}\mathcal{C}%
_{III}^{l}.
\end{equation}%
$M_{I\rightarrow II}$ is the transfer matrix from the region before the barrier to within the barrier and $M_{II\rightarrow III}$ is the one from within to behind the barrier. Using the matrix form of the spinor wave functions in Eq. (\ref{eq:MatrixSpinor}), the transfer matrix is given by
\begin{equation}
M_{I\rightarrow II} = \mathcal{E}^{-1}_I \mathcal{P}^{-1}_I \mathcal{P}_{II} \mathcal{E}_{II}.
\end{equation}
Finally the transmission $\left( T\right) $ and reflection $\left( R\right) $ probabilities are obtained as
\begin{equation}
T_{\pm }^{l}=\frac{k^{\pm }}{k^{l}}\left\vert t_{\pm }^{l}\right\vert ^{2}%
\text{ and }R_{\pm }^{l}=\frac{k^{\pm }}{k^{l}}\left\vert r_{\pm
}^{l}\right\vert ^{2}.  \label{TProb}
\end{equation}%
This takes into account the change in velocity of the waves when they are scattered into a different propagation mode.

Some of the different probabilities in Eq. $\left( \ref{TProb}\right) $ can be related to each other via the time reversal symmetry of the system. The Hamiltonian under consideration describes electrons in the vicinity of one of the two Dirac points in reciprocal space which are called the two \textit{valleys}. The Hamiltonian $H^{\prime }$ for the other Dirac point is given by
\begin{equation}
H^{\prime }=-H^{T}.
\end{equation}%
This means that electrons scattering from $k^{+}$ to $k^{-}$ when moving from left to right in the first valley are equivalent to electrons scattering from $k^{-}$ to $k^{+}$ but moving in the opposite direction and in the other valley. When $\delta=0$, both valleys are equivalent and time reversal symmetry holds near a single Dirac point. Therefore, the transmission probability of electrons moving in the opposite direction must be the same because of the valley equivalence. Using a similar argument for the reflection in both valleys, one can conclude that
\begin{equation}
T_{-}^{+}=T_{+}^{-}\text{ and }R_{-}^{+}=R_{+}^{-}.
\end{equation}
Another symmetry operation ensures the symmetry of the probabilities with respect to normal incidence. Note that the Hamiltonian of the system given in Eq. (\ref{StartHam}) is not symmetric under a sign flip in $k_y$. An interchange of the $A_1$ and $B_2$ atoms together with $B_1 \leftrightarrow A_2$ however leaves the system invariant, but corresponds to exchanging $k_y \rightarrow -k_y$. Since the system is invariant under this transformation, he transmission and reflection should be symmetric with respect to normal incidence. The application of an interlayer bias however breaks this exchange symmetry and therefore asymmetric results in the transmission and reflection are expected. This asymmetry was noted before by Nilsson \textit{et al.}\cite{Nilsson2007}. The application of an interlayer bias furthermore lifts the valley degeneracy and with it the above discussed symmetry in the scattered transmission. Note that when the same quantities are calculated for states in the other valley, the asymmetry is reversed and therefore the overall symmetry of the system is preserved.

In the two band model, one obtains a two spinor which can be described by a similar matrix product as before with
\begin{equation}
\mathcal{P}_{2}=\left[
\begin{array}{cccc}
1 & 1 & 1 & 1 \\
-\frac{\left( k-ik_{y}\right) ^{2}}{\Gamma_{1}\varepsilon } & -\frac{\left(
k+ik_{y}\right) ^{2}}{\Gamma_{1}\varepsilon } & \frac{\left( \kappa
+k_{y}\right) ^{2}}{\Gamma_{1}\varepsilon } & \frac{\left( \kappa -k_{y}\right)
^{2}}{\Gamma_{1}\varepsilon }%
\end{array}%
\right] ,
\end{equation}%
where $k$ and $\kappa $ are defines as before. Equating both wave functions and their derivatives at the edges of the barrier leads to the transmission probability $T$.

\section{Numerical results}\label{NumResults}

\begin{figure}[tb]
\begin{center}
\includegraphics[width=8cm]{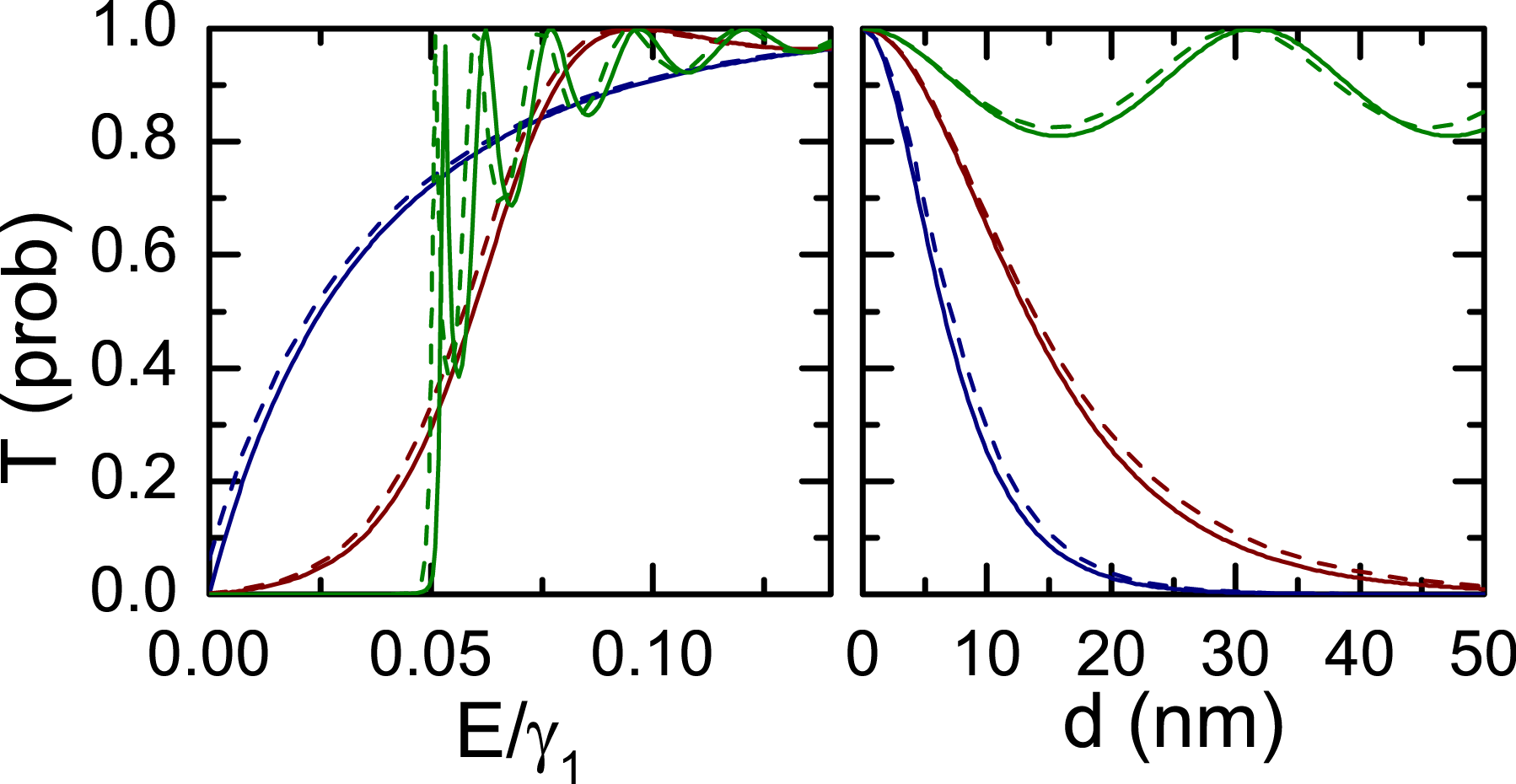}
\end{center}
\caption{(Colour online) Comparison of the transmission probability through a barrier of height $V_0 = 0.05 \gamma_1$ with (dashed curves) and without (solid curves) the skew hopping parameters. (Left) The energy dependence of the transmission probability for junctions of width $d=10 nm$ (blue), $d=25 nm$ (red) and $d=100 nm$ (green). (Right) The width dependence of the transmission probability for a Fermi energy of $E=\frac{1}{5}V_0$ (blue), $E=\frac{2}{5}V_0$ (red) and $E=\frac{8}{5}V_0$. }
\label{Fig:TransPerp}
\end{figure}

In order to investigate the importance of the skew hopping parameters, we present in Fig. \ref{Fig:TransPerp} the transmission probability through a potential barrier at normal incidence with and without including the skew parameters as function of the Fermi energy and the width of the barrier. The results show that even in the low energy range where the effect of the skew hopping parameters are expected to be the largest, the transmission probabilities are very similar. We therefore conclude that for the discussion at hand these parameters are not important and we will neglect them in the following discussion.

\begin{figure*}[th]
\begin{center}
\includegraphics[width=14cm]{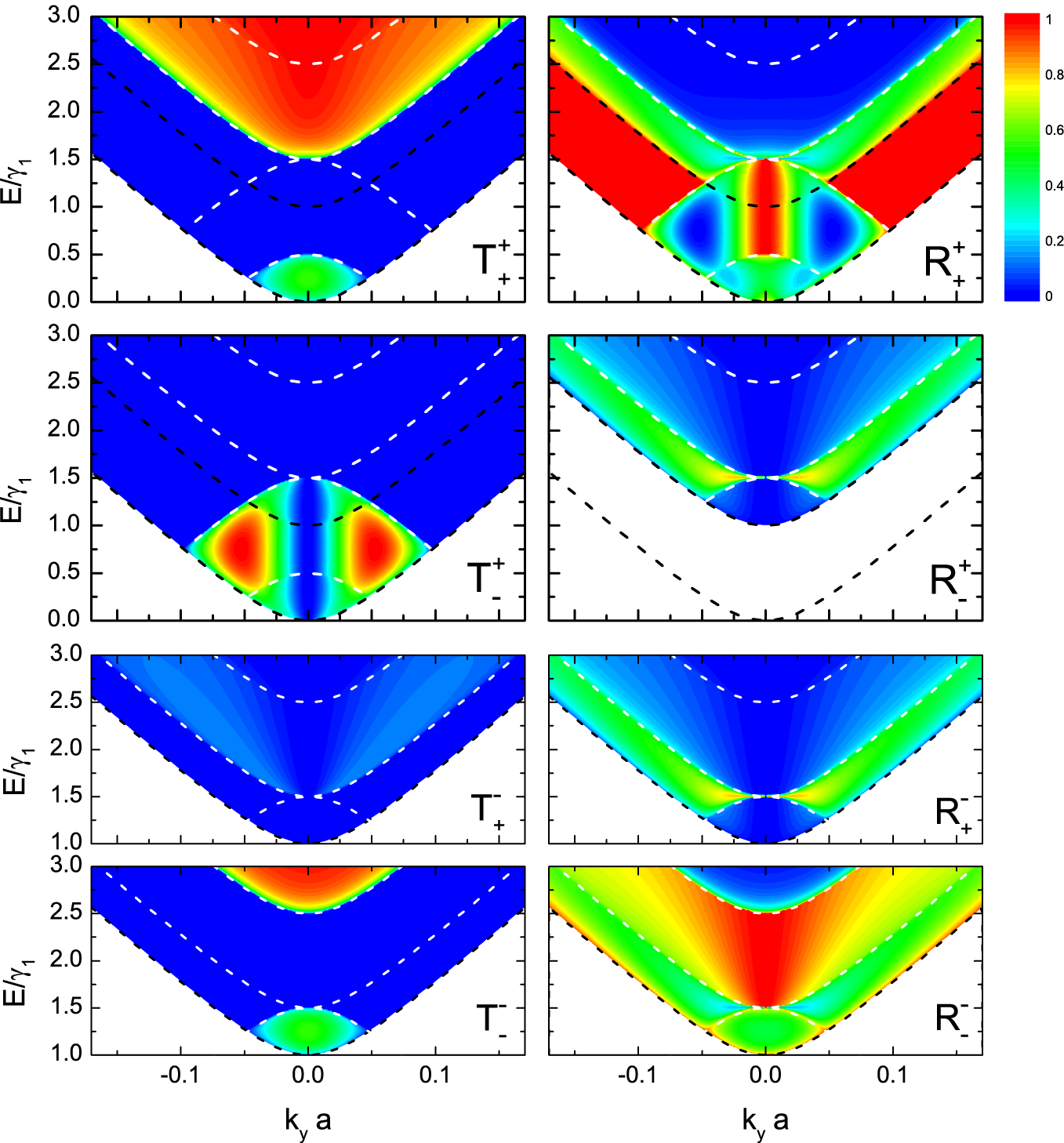}
\end{center}
\caption{(Colour online) The transmission and reflection probabilities through a pn-junction of height $V_{0}=\frac{3}{2}\protect\gamma _{1}$ as function of energy and transverse wave vector $k_{y}$. The energy is expressed in units of $\protect\gamma _{1}$ and the wave vector in units of $a^{-1}$, the inverse of the in plane interatomic distance. The dashed curves indicate the borders between the regions where different modes are propagating or are evanescent inside or outside the junction.}
\label{Step}
\end{figure*}

In Fig. \ref{Step} we show the transmission and reflection probabilities for a pn-junction as a function of the energy of the incident wave $E$ and its transverse wave vector $k_{y}$. The height of the potential is set to $V_{0}=\frac{3}{2}\gamma _{1}$ and the interlayer potential difference $\delta=0$. The results show qualitatively different regions in the $\left( k_{y},E\right) $-plane which can be explained by identifying which modes are propagating inside and outside the pn-junction. The borders between these regions are indicated by dashed curves superimposed on the density plots.

For normal incidence, $k_{y}=0$, the expected cloaking in the $T_{+}^{+}$ and $T_{-}^{+}$ channels occurs for $V_{0}-\gamma _{1}<E<V_{0}$. When the transverse wave vector $k_{y}$ differs from zero, it is possible for electrons to scatter into the $k^{-}$ propagating mode inside the pn-junction and this results in a scattered transmission in the $T_{-}^{+}$ channel. For energies smaller than $V_{0}-\gamma _{1}$, there are propagating $k^{+}$-states in the junction and this leads to a non zero transmission in the $T_{+}^{+}$ channel. This transmission is absent when the potential $V_{0}<\gamma _{1}$ and it is also not present in the two band model. For energies larger than the height of the barrier, $E>V_{0}$, the particles behave similar to Schr\"{o}dinger particles.

The numerical results of the probabilities $T_{-}^{-}$ and $R_{-}^{-}$ are shown in the fourth row of Fig. \ref{Step}. For $E<V_{0}$ the electrons tunnel using the propagating $k^{-}$ states in the junction since the states match the number $l=-1$. For $V_{0}<E<V_{0}+\gamma _{1}$ however there are no available $k^{-}$ states and the transmission is suppressed even though the energy is larger than the height of the potential barrier. This is the equivalent phenomenon of the \textit{cloaking} discussed earlier. This phenomenon is absent in the two band approximation.

The scattered reflection $\left( R_{-}^{+},R_{+}^{-}\right) $ shown in the second and third row of Fig. \ref{Step} is very strong at near normal incidence and $E\approx V_{0}$. In this region both states are propagating outside the junction, but are evanescent inside it. In this situation scattering is favorable due to the symmetry of the scattering probabilities. This argument is independent of the mode the incident wave is in. For a pn-junction the symmetry argument of the previous section only holds for the scattered reflection. Because inside the junction the carriers behave as hole states while outside they are electrons, the symmetry described in Sec. \ref{TimeSymmetry} is broken and therefore the equivalence in scattered transmission is no longer valid. Since under reflection the electrons return again in an electron state, for the reflection channel the symmetry remains valid, which is also seen in the calculations of the reflection channels $R_{-}^{+}$ and $R_{+}^{-}$.

\begin{figure*}[th]
\begin{center}
\includegraphics[width=14cm]{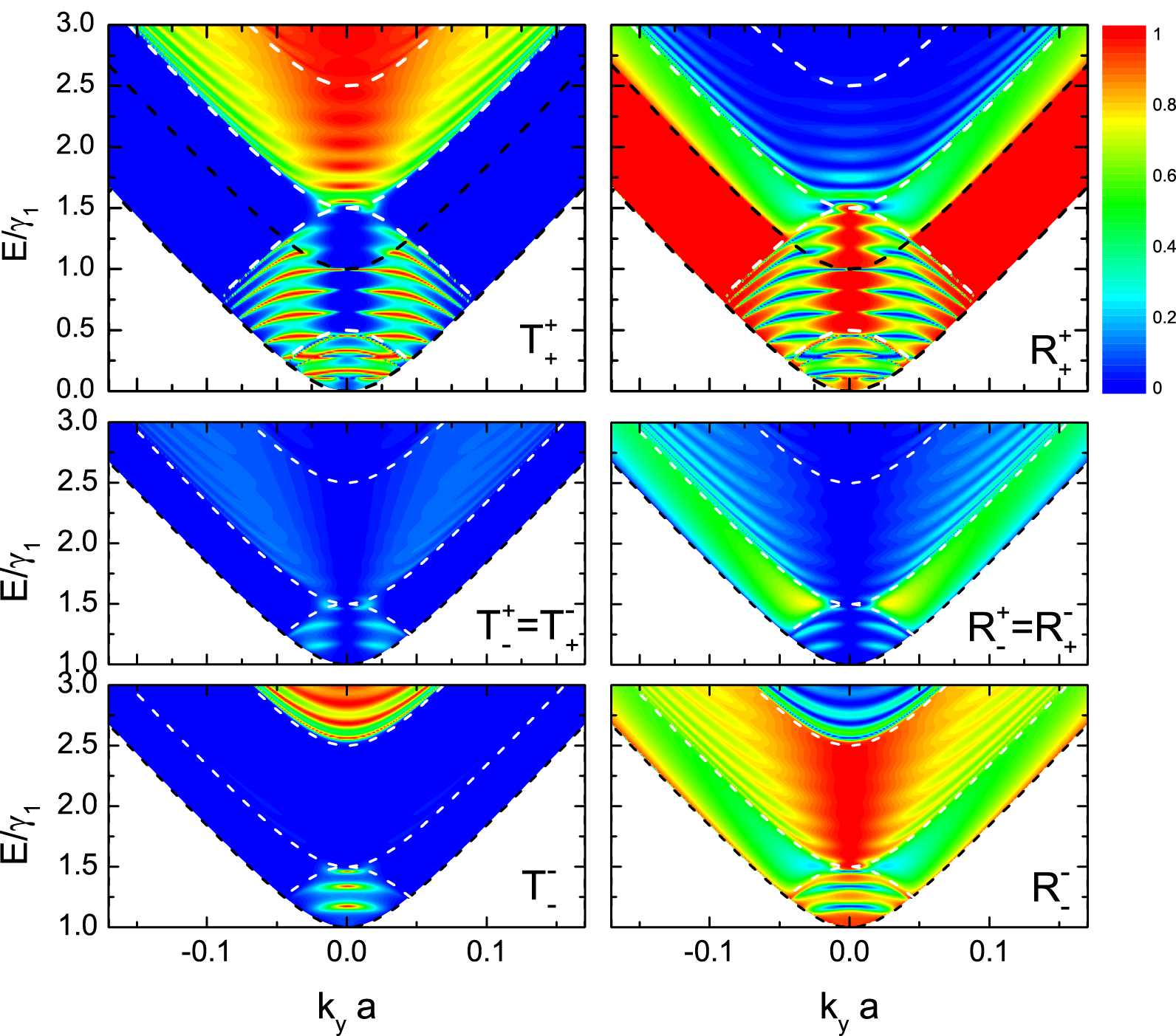}
\end{center}
\caption{(Colour online) The same as in Fig. \ref{Step}, but now for the transmission and reflection probabilities through a potential barrier of height $V_{0}=\frac{3}{2}\protect\gamma _{1}$ and width $d=25$nm.}
\label{PPcombi}
\end{figure*}

In Fig. \ref{PPcombi} we show the transmission and reflection probabilities for a potential barrier of width 25nm and the same height as the pn-junction. The cloaking in the $T_{+}^{+}$ and $T_{-}^{-}$ channel occurs for the same conditions as the pn-junction. When the transverse wave vector $k_{y}$ differs from zero, it is possible for electrons to scatter into the $k^{-}$ propagating mode inside the barrier and this results in the observation of resonances which follow the expression given by Eq. $\left( \ref {eq:Resonances}\right) $ for $k_{y}$ large. For energies smaller than $V_{0}-\gamma _{1}$, there are propagating $k^{+}$-states in the barrier and resonances appear which follow the expression given by Eq. $\left( \ref{eq:Resonances}\right) $. These resonances are absent when the potential $V_{0}<\gamma _{1}$ and are also not present in the two band model. The symmetry argument of the previous section does hold for a barrier since at both sides of the barrier the particles have positive energy. Therefore the scattered transmission is the same as shown in the second row of Fig. \ref{PPcombi}.

\begin{figure*}[th]
\begin{center}
\includegraphics[width=14cm]{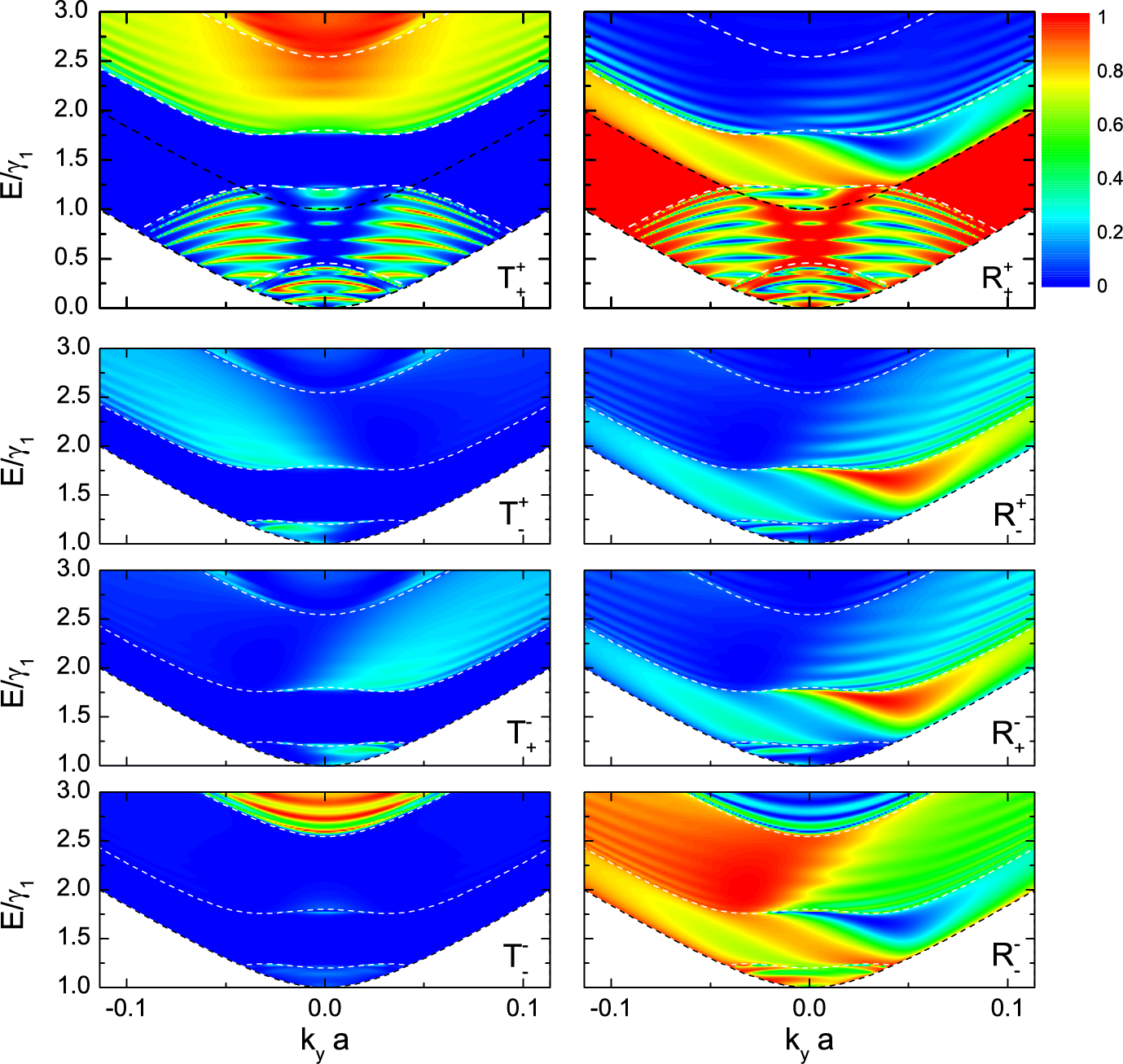}
\end{center}
\caption{(Colour online) The same as in Fig. \ref{Step}, but now for the transmission and reflection probabilities through a biased potential barrier of height $V_{0}=\frac{3}{2}\protect\gamma _{1}$ and width $d=25$nm and interlayer bias $\delta = 0.3\gamma_1$.}
\label{DeltaPlots}
\end{figure*}

In Fig. \ref{DeltaPlots} the transmission and reflection probabilities for a barrier of height $V_0=\frac{3}{2} \gamma_1$ and interlayer bias $\delta = 0.3\gamma_1$ are plotted. The bandgap introduced by the interlayer bias suppresses the transmission in the energy region between $V_0 \pm \delta$. In the bandgap a remarkable asymmetric feature with respect to normal incidence in the reflection channels shows up. This is a manifestation of the breaking of the interlayer sublattice equivalence as discussed in Sec. \ref{TimeSymmetry}. This asymmetry is also present in the scattered transmission where it depends on the incident mode. This asymmetric feature is only present when scattering between different propagation modes is possible. It is therefore an other qualitative feature that is not present in the two band approximation.

The interlayer potential difference furthermore couples the two propagation modes at normal incidence. Therefore the suppression due to cloaking is also adjusted. Now both the interlayer bias term $\delta$ as the transverse momentum $k_y$ causes the two modes to be coupled as shown in Eq. (\ref{eq:HamTransfo}). Due to the interplay of both parameters, cloaking at normal incidence splits into two branches at finite $k_y$ as shown in the non scattered transmission and reflection.

\begin{figure}[th]
\begin{center}
\includegraphics[width=8cm]{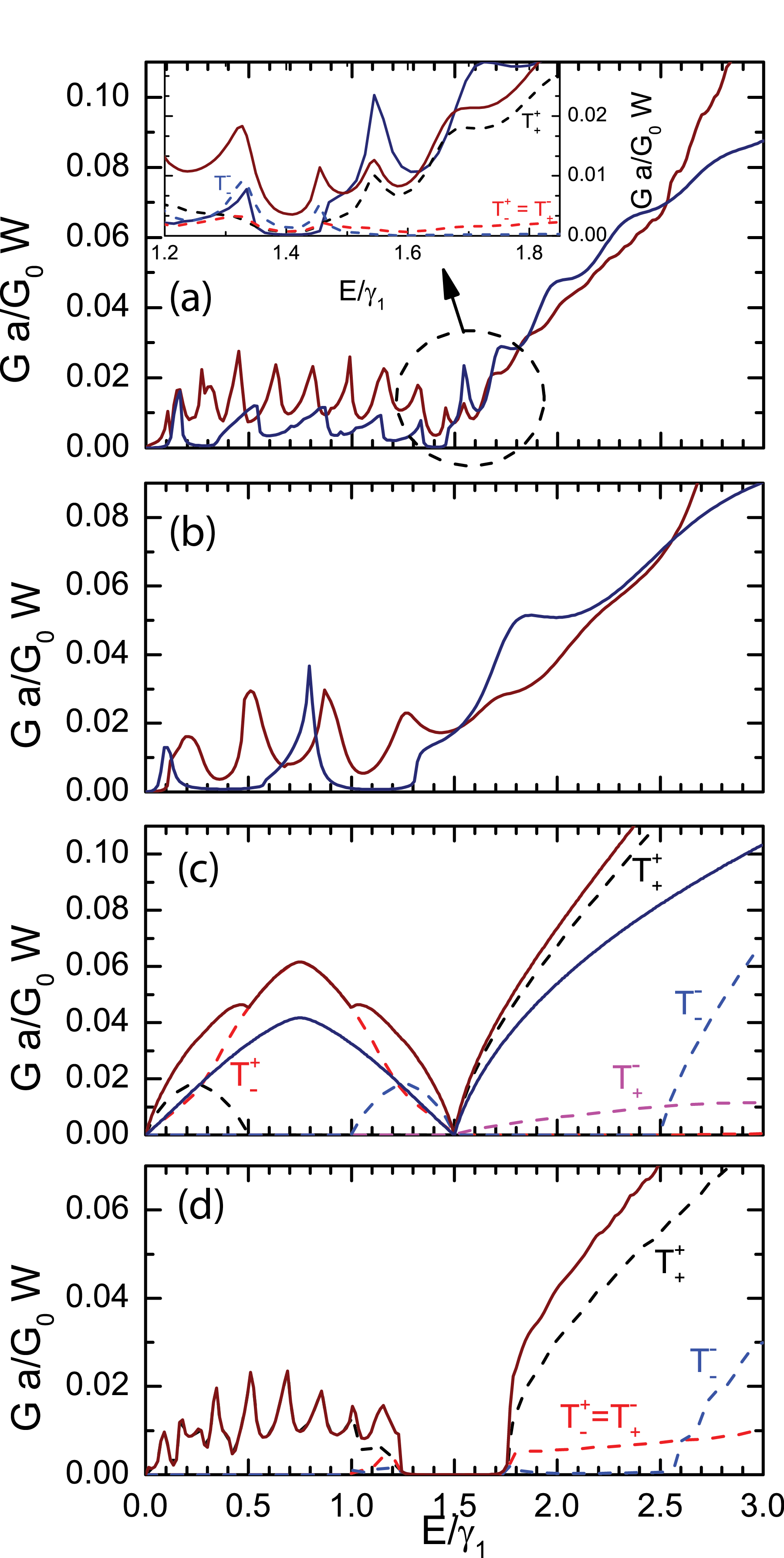}
\end{center}
\caption{(Colour online) Energy dependence of the conductance of a single barrier with $d=25nm$ (a), $d=10nm$ (b) and a pn-junction with height $V_{0}= \frac{3}{2}\protect\gamma _{1}$ calculated using the four band method (blue) and the two band approximation (dark red). Figure (c), (d) and the inset of (a) show the different contributions of the four transmission channels as dashed lines. (d) Conductance for a biassed bilayer with interlayer bias $\delta=0.3\gamma_1$ and potential barrier the same width and height as that of (a).}
\label{ConductancePlot}
\end{figure}

The conductance is shown in Fig. \ref{ConductancePlot} for both the two band approximation and the four band method for barriers of different width (\ref{ConductancePlot}(a) and \ref{ConductancePlot}(b)) and a pn-junction (\ref{ConductancePlot}(c)). For energies smaller than the barrier's height, $E<V_{0}$, the resonances in the transmission show up as peaks in the conductance. Using the four band method those resonances are however more pronounced and there is a difference in the position and number of peaks. Furthermore for energies smaller than $V_{0}-\gamma _{1}$, the resonance peaks for the propagation via the $k^{+}$-states appear as shoulders of the other peaks. This phenomenon does not occur in the two band approximation for which all the resonances are similar. When the energy is larger than $\gamma _{1}$, additional peaks result from propagation via the $k^{-}$-states inside the barrier while peaks of the two band approximation do not differ. This is clarified in the inset of Fig. \ref{ConductancePlot}(a) showing the contributions of the different transmission channels. While the contribution of the $T_{+}^{+}$ channel is low in this region, the $T_{-}^{-}$ and the scattered transmission channels have well pronounced resonances. When the energy is larger than the height of the barrier, $E>V_{0}$, the conductance predicted by both models is of the same order of magnitude. However, when the energy is larger than $V_{0}+\gamma _{1}$, the $k^{-}$-state is not cloaked anymore resulting in additional conductance that is absent for the two band model.

Snyman \textit{et al. }\cite{Snyman2007} have shown that both models coincide in the region for low barrier height, i.e. $V_{0}<\gamma _{1}$, which is confirmed by our calculations. The characteristic form of the conductance resembles that of the result obtained here for the two band approximation. Now it is however clear that the conductance peak just above the barrier height is lower in the four band model. This is due to the additional scattered reflection channel $R_{-}^{+}/R_{+}^{-}$ that is absent for low barriers. In the latter case only the $T_{+}^{+}$ channel could contribute and coincides with the two band model transmission, but now this channel is suppressed.

Fig. \ref{ConductancePlot}(b) shows that the resonant peaks in the conductance depend on the width of the barrier but note that they occur at different energies in both models. In Fig. \ref{ConductancePlot}(c), the conductance of a PN junction is calculated. Although the barrier is of infinite width, both models predict a finite conductance. For the four band model however, the different energy regions show up as bumps in the conductance caused by the availability of the second mode of propagation.

Fig. \ref{ConductancePlot}(d) shows the conductance for a biased potential barrier. The results are similar to that of the unbiased case but are influenced by the suppression of the conductance in the energy range of the bandgap at $V_0 \pm \delta$. Note that although the scattered transmission is asymmetric with respect to normal incidence, this is not visible in the conductance calculations since it sums over all values of $k_y$.

\section{Conclusion}\label{Conclusion}

We evaluated the transmission and reflection of electrons through potential barriers and a pn-junction in bilayer graphene. We extended previous calculations performed within the two band model to the four band model. We compared the results with and without taking into account the skew hopping parameters and found that the latter can be neglected for energy ranges $E>\gamma_1/100\approx 4meV$. Within the four band model, the results from the two band approximation are recovered for small energies and low potential barriers. We find new phenomena such as transmission resonances at normal incidence. We showed that the notion of pseudospin used to describe electrons in the two band approximation corresponds to the wavefunctions being symmetric or antisymmetric with respect to in-plane mirroring and that this leads to the observation of \textit{cloaking} which occurs both for symmetric and antisymmetric states. We have also pointed out that because it is a consequence of the symmetry of the system, the notion of pseudospin also holds when the skew hopping parameters are taken into account. For high energies a new mode of propagation is available for the electrons, which is not present in the two band approximation, and we found that for non normal incidence it is possible to scatter between the two modes. The resulting conductance of the four band model incorporates these new phenomena and therefore differs significantly from the conductance calculated within the two band model. This difference manifests itself by the presence of many more and well defined resonances and a substantially higher conductance for high energies. Finally, we showed that the application of an interlayer bias in the system significantly changes the transmission characteristics. The bandgap created by the interlayer potential forms a distinct feature in the conductance and the transmission and reflection probabilities lack reflection symmetry due to normal incidence. The introduction of the interlayer symmetry breaking term furthermore couples the symmetric and antisymmetric modes. Therefore the notion of pseudospin is no longer valid.

\section{Acknowledgements}

This work was supported by the European Science Foundation (ESF) under the EUROCORES Program Euro-GRAPHENE within the project CONGRAN, the Flemish Science Foundation (FWO-Vl) and the Methusalem Programme of the Flemish Government.


\begin{thebibliography}{99}
\bibitem{Bernal1924} J.~D. Bernal, \newblock Proc. R. Soc. A, \textbf{106},
749 (1924).

\bibitem{McCann2006} E.~McCann and V.~I. Fal'ko, \newblock Phys. Rev. Lett. \textbf{96}, 086805 (2006).

\bibitem{Novoselov2006} K.~S. Novoselov, E. McCann, S. V. Morozov, V. I. Fal\'ko, M. I. Katsnelson, U. Zeitler, D. Jiang, F. Schedin, and A. Geim,
\newblock Nat. Phys. \textbf{2}, 177 (2006).

\bibitem{Partoens2006} B.~Partoens and F.~M. Peeters, \newblock Phys. Rev. B \textbf{74}, 075404 (2006).

\bibitem{McCann2006d} E. McCann, \newblock Phys. Rev. B \textbf{74}, 161403 (2006).

\bibitem{Zhang2009} Y. Zhang, T. T. Tang, C. Girit, Z. Hao, M.C. Martin, A. Zettl, M.F. Crommie, Y.R. Shen, and F. Wang, \newblock Nature \textbf{459}, 820 (2009).

\bibitem{Castro2007} E. V. Castro, K. S. Novoselov, S. V. Morozov, N. M. R. Peres, J. M. B. Lopes dos Santos, J. Nilsson, F. Guinea, A. K. Geim, and A. H. Neto, \newblock Phys. Rev. Lett. \textbf{99}, 216802 (2007).

\bibitem{Blanter2000} Ya.~M. Blanter and M.~B\"{u}ttiker, \newblock Phys. Rep.
\textbf{336}, 1 (2000).

\bibitem{Katsnelson2006} M.~I. Katsnelson, K.~S. Novoselov, and A.~K. Geim,
\newblock Nat. Phys. \textbf{2}, 620 (2006).

\bibitem{Young2009} A.~F. Young and P.~Kim, \newblock Nat. Phys. \textbf{5},
222 (2009).

\bibitem{Stander2009} N.~Stander, B.~Huard, and D.~Goldhaber-Gordon, %
\newblock Phys. Rev. Lett. \textbf{102}, 026807 (2009).

\bibitem{Barbier2010} M.~Barbier, P.~Vasilopoulos, and F.~M. Peeters, %
\newblock Phys. Rev. B \textbf{82}, 235408 (2010).

\bibitem{Gu2011} N.~Gu, M.~Rudner, and L.~Levitov, \newblock Phys. Rev.
Lett. \textbf{107}, 156603 (2011).

\bibitem{Efetov2010} D.K. Efetov and P. Kim,\newblock Phys. Rev. Lett. \textbf{105}, 256805 (2010).

\bibitem{Ye2011} J. Ye, M.F. Craciun, M. Koshino, S. Russo, S. Inoue, H. Yuan, H. Shimotani, A.F. Morpurgo, and Y. Iwasa, \newblock Proc. Natl. Acad. Sci. USA \textbf{108}, 13002 (2011).

\bibitem{Efetov2011} D.K. Efetov, P. Maher, S. Glinskis, and P. Kim, \newblock Phys. Rev. B \textbf{84}, 161412 (2011).

\bibitem{Partoens2007} B.~Partoens and F.~M. Peeters, \newblock Phys. Rev. B
\textbf{75}, 193402 (2007).

\bibitem{McCann2007} E.~McCann, D.~S.~L.~ Abergel, and V.~I. Fal'ko, \newblock Solid State Commun.
\textbf{75}, 193402 (2007).

\bibitem{Snyman2007} I.~Snyman and C.~W.~J. Beenakker, \newblock Phys. Rev.
B \textbf{75}, 045322 (2007).

\bibitem{Neto2009} A.~Neto, F.~Guinea, and N.~Peres, \newblock Rev. Mod.
Phys. \textbf{81}, 109 (2009).

\bibitem{Matulis2009} A.~Matulis and F.~M. Peeters, \newblock Am. J. Phys.
\textbf{77}, 595 (2009).

\bibitem{Nilsson2007} J. Nilsson, A.H. Castro Neto, F. Guinea, and N. M. R. Peres, \newblock Phys. Rev. B \textbf{76}, 165416 (2007).

\end{thebibliography}
\end{document}